\documentstyle[11pt,newpasp,twoside,epsf]{article}
\def\oiii{[O~{\sc iii}]}
\def\ha{H$\alpha$}
\def\stry{{Str\"omgren $y$}}

\def\sii{[S~{\sc ii}]}
\def\heii{He~{\sc ii}}
\def\ifilter{{$i^\prime$}}
\def\gfilter{{$g^\prime$}}
\def\rfilter{{$r^\prime$}}

\begin{document}
\title{A search for Symbiotic Stars in the Local Group}
 \author{Laura Magrini}
\affil{Dipartimento di Astronomia e Scienza dello Spazio, Largo E. Fermi 2, I-50125 Firenze. Italy}
 \author{Romano L. M. Corradi}
\affil{Isaac Newton Group, Apartado de Correos 321, 38700 Santa Cruz de la Palma-Spain} 
 \author{Ulisse Munari}
\affil{Osservatorio Astronomico di Padova, Sede di Asiago, I-36012 Asiago (VI), Italy}

\begin{abstract}
The Local Group Census is a narrow- and broad-band survey of all the
galaxies of the Local Group above $\delta =-30^\circ$, in progress at the
2.5m Isaac Newton telescope on La Palma.  We discuss here the ability of the
survey to detect symbiotic star candidates in the Local Group, by deriving
detection limits in each of the narrow- and broad-band frames used in the
survey, and by estimating the total number of objects expected in each
galaxy. We present two diagnostic diagrams, based on the adopted photometric
filters, to discriminate between symbiotic stars and other emission-line
objects such as planetary nebulae.
\end{abstract}

\section{Introduction}

The Local Group Census (LGC) is a narrow- and broad-band survey of all the
galaxies of the Local Group (LG) above $\delta =-30^\circ$
(http://www.ing.iac.es/ WFS/LGC/).  It is carried out as part of the Isaac
Newton Group's Wide Field Survey program and its observations are being
obtained with the Wide Field Camera at the 2.5m Isaac Newton telescope,
equipped with a 4-CCD mosaic covering a field of view of 
$34^\prime \times 34^\prime$.\\

The LG galaxies have been observed through narrow-band filters (\oiii\
$\lambda$5007\AA, \ha, \heii\ $\lambda$4686\AA, and \sii\ $\lambda$6725\AA)
and broad band filters (Sloan \gfilter, \rfilter, \ifilter, and \stry).  One
of the goals of the LGC is the search for symbiotic star candidates in
galactic systems other than the Milky Way, using the broad-band color index
\gfilter --\ifilter\ to detect the red giants and the narrow-band \ha\
filter to select those with a strong emission from circumbinary gas.

\section{Diagnostic diagrams to identify symbiotic stars}

The images taken for the Local Group Census are quite deep and
complete including observations of most of the northern LG galaxies.
Our aim is to find simple diagnostic diagrams allowing to detect
symbiotic star candidates using only the information provided by our
survey.  The nebulae around symbiotic systems are often similar to
PNe in terms of excitation conditions and chemical abundances, in
particular those around symbiotic Miras.  The spectrum of some of the
latter in the region 4000 to 7000 \AA\ is in fact quite
indistinguishable from that of planetary nebulae.

Using data from the recent multi-epoch spectrophotometric atlas of
symbiotic stars by Munari \& Zwitter (2002), we have built and
evaluated two diagnostic diagrams which should be useful in
distinguishing between symbiotic stars and PNe on the base of the
photometric filters used in our survey. The presence of a cool giant
is easily revealed by comparing \ifilter\ and \stry, the circumstellar
material stands up well by comparing \ha\ and \stry\, and the high
electronic density conditions characterizing symbiotic stars affects
the relation between \oiii\ and \stry.

\begin{figure}
\plottwo{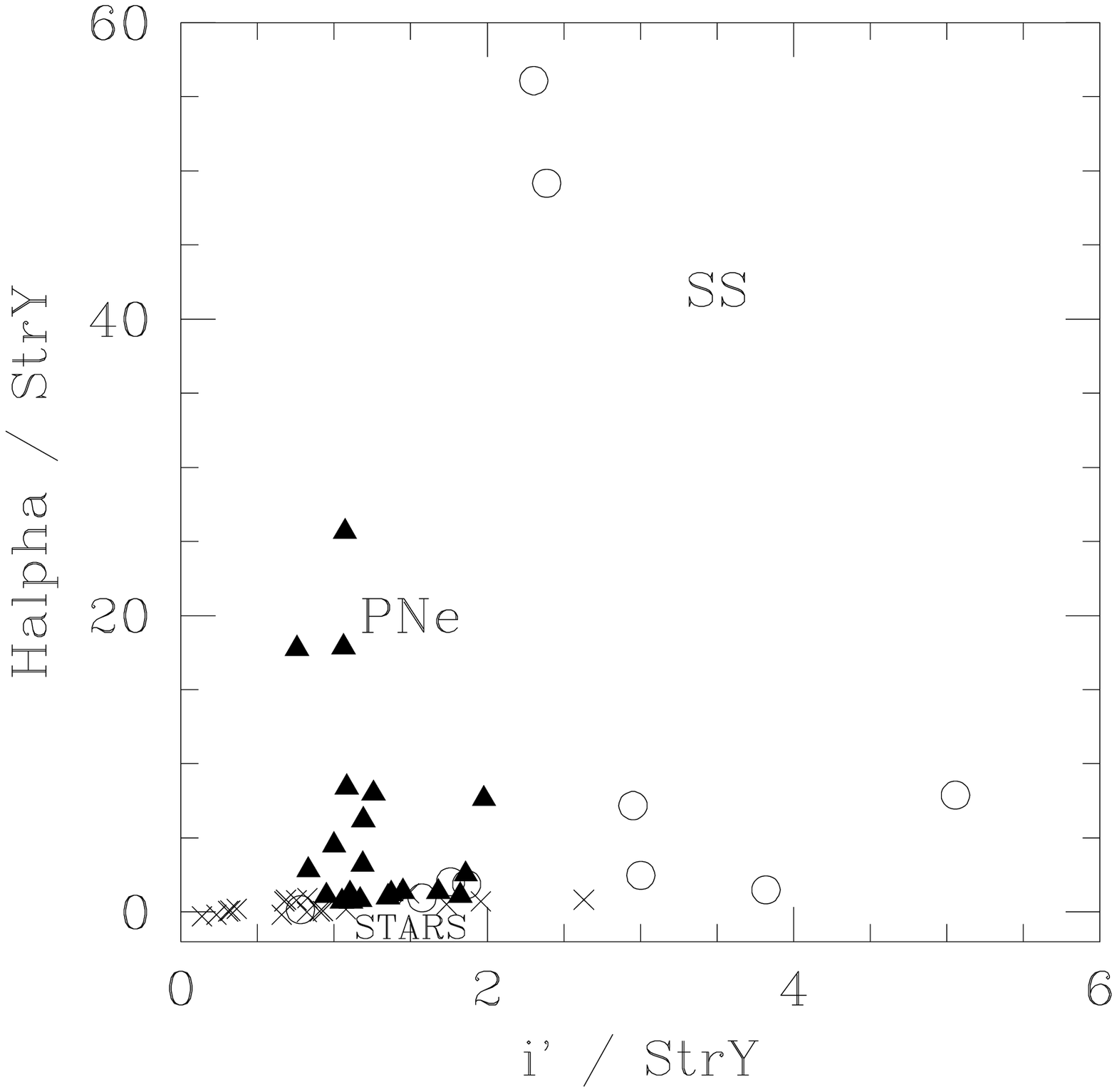}{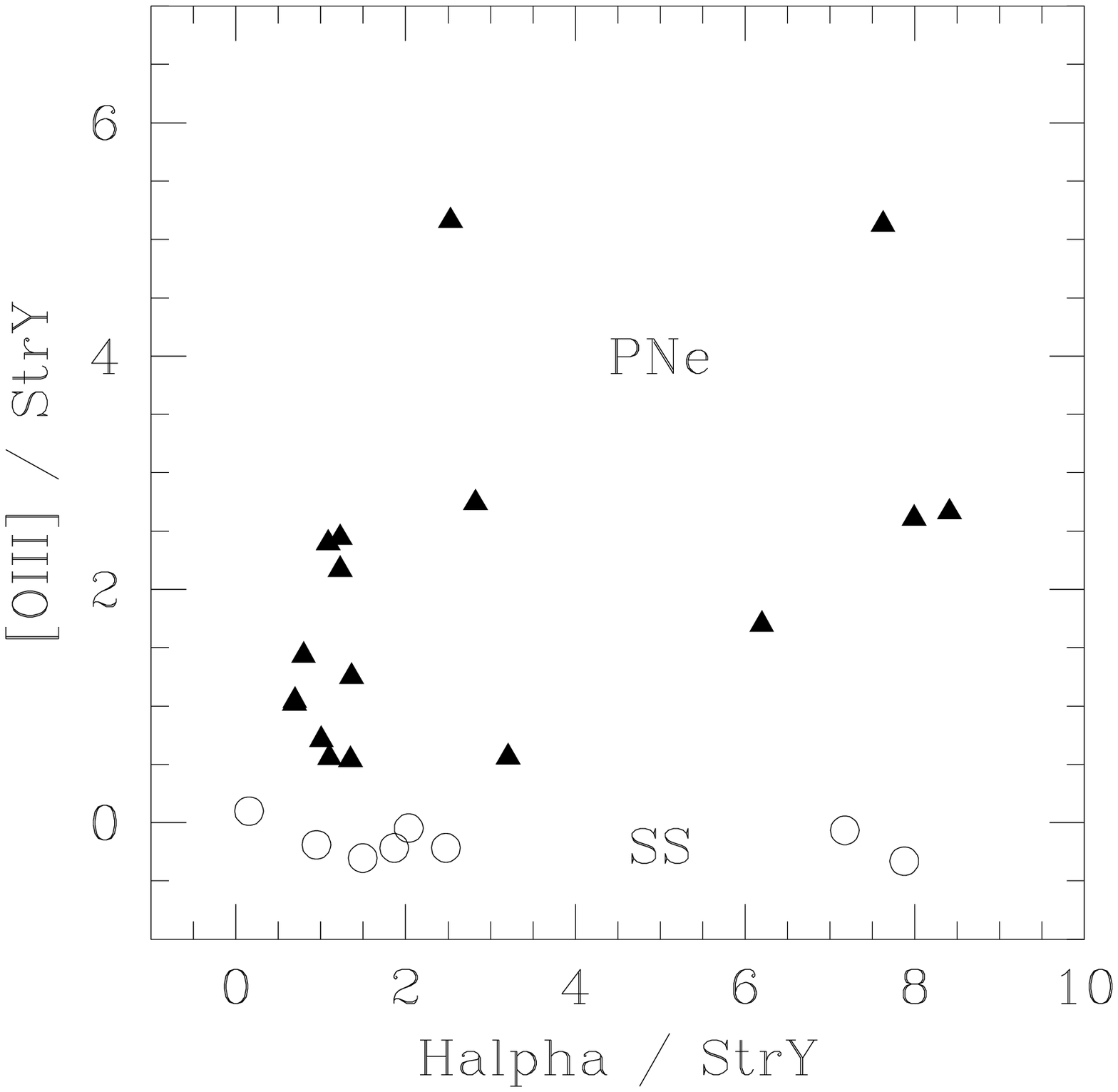}
\caption{Diagnostic diagrams useful in the search for extra-galactic 
symbiotic stars. Triangles are PNe in M33 discovered by
Magrini et al. (2000, 2001), empty circles are symbiotic stars
from Munari \& Zwitter (2002) spectrophotometric atlas, 
and crosses are normal field stars.}
\end{figure}

\begin{table}[!t]
\begin{center}
{\footnotesize
\begin{tabular}{lllll} \hline
Name     & Type    	& M 		&N.Symbiotic Stars	\\
               &                	& (M$_{\odot}$)	& ($K-B$)\\
\hline 
Milky Way & Sb I-II	&1.8-3.7$\times$10$^{11}$		&400,000\\
WLM	& IrrIV-V   & 1.5$\times$10$^8$		&0\\
IC 10     	& IrrIV     	&6.$\times$10$^8$			&150\\
NGC 147   & Sph  	&5.5$\times$10$^7$		&2,800\\
NGC 185   & Sph 	&6.6$\times$10$^8$		&4,200\\
NGC 205   & Sph 	&7.5$\times$10$^8$		&17,000\\
M 32      	& E2        	&1.1$\times$10$^9$		&19,000\\
M 31      	& SbI-II    	&2-4$\times$10$^{11}$		&660,000\\
LGS 3     	& dIrr/dSph &2.6$\times$10$^7$		&-\\
IC 1613   	& IrrV     	&1.$\times$10$^8$			&0\\
M 33      	& ScII-III  	&0.8-1.4$\times$10$^{10}$		&45,000\\
Fornax    	& dSph      	&6.8$\times$10$^7$	 	&500\\
EGB0427$+$63& Ir    &    				&-\\
Leo A     	& Ir V    	&$<$9.$\times$10$^7$		&0\\
Sextans B 	& IrIV-V    &8.8$\times$10$^8$		&0\\
NGC 3109 & IrIV  	&2.3$\times$10$^9$		&0\\
Leo I           & dSph      &$>$2.$\times$10$^7$		&200\\
Sextans A 	& IrV      	&$>$1.9$\times$10$^8$		&0\\
Leo II          & dSph	&1.1$\times$10$^7$ 		&50\\
GR 8      	& dIrr      	&				&0\\
Ursa Minor& dSph      &1.7$\times$10$^7$		&0\\
Draco     	& dSph     	&1.7$\times$10$^7$		&10\\
Sagittarius& dSph  	&1.5$\times$10$^8$		&-\\
NGC 6822 & IrrIV-V  &1.7$\times$10$^9$		&0\\
DDO 210   & dIrr      	&1.4$\times$10$^7$		&0\\
Pegasus   	& IrV 	&2.7$\times$10$^7$		&0\\ 
\hline 
\end{tabular}
\caption{Galaxies in the LGC program and the expected number of symbiotic systems
computed from $K$ and $B$ data (see sec.4 for details). Only a small part of them could be in
an active state (bright emission lines) and therefore been detectable by our survey.}}
\end{center}
\end{table}

\section{ How far can a symbiotic star be detected in our Survey?}

To estimate the sensitivity in distance of our survey, we have picked
out several prototype symbiotic stars from the atlas of Munari \&
Zwitter (2002), namely: carbon symbiotic stars, symbiotic novae in
outburst, classical symbiotic stars in outburst, quiescent yellow and
normal symbiotics, symbiotic Miras.  With filter band profiles taken
from Moro \& Munari (2000), we have derived via synthetic photometry
the flux in each filter and corrected for the symbiotic star distance
estimated via infrared spectrophotometric parallax of the cool
giant. We did not correct for reddening to account 
for typical reddening conditions encountered in the LG.
The scaled fluxes (\ha, \oiii, \stry\, and \ifilter) of the stars in the
sample span a relatively narrow range, and this allows to consider the
median value of the fluxes in each filter in computing the greatest distance
to detect a typical symbiotic star with the LGC survey.

Using the WFC at the prime focus of the Isaac Newton telescope (2.5m) and
the exposure times set by the LGC (3600 s through the \ha\ and \oiii\
filters, 1800 s through the \stry\ filter and 1200 s through the \ifilter\
filter), a typical symbiotic star can be observed with a signal to noise of
5 at the distance of approximately 0.5 Mpc through the \ha\ filter, of 0.2
Mpc through the \oiii\ filter, 0.1 Mpc through the \stry\ filter and 0.6 Mpc
through the \ifilter.  The brightest symbiotic star of our sample can be
observed at the distance of 1.6 Mpc (\ha), of 1.4 Mpc (\oiii), of 0.8 Mpc
(\stry) and of 1.6 Mpc (\ifilter). These limits suggest that in the closer
LG galaxies we can detect symbiotic stars over a large range of intrinsic
luminosity, while for the farthest LG sample only the brightest objects will
be reveal with a useful S/N ratio.

\section{How many symbiotic stars could be detected in each LC galaxy?}

Few symbiotic stars are known in external galaxies: 8 in the LMC, 6 in the
SMC and 1 in Draco (Belczy\'nski et al. 2000).  No one of these surveys is
complete and so they cannot give information about the number of the
expected symbiotic stars in each LG galaxy.  A total number of $3 \times
10^5$ symbiotic star is estimated in our Galaxy by Munari \& Renzini (1992),
$3 \times 10^4$ by Kenyon et al. (1992) and $3 \times 10^3$ by Allen (1984).
So far only $\sim 200$ have been discovered.

We considered a qualitative way to estimate the approximate expected number
of symbiotic stars in each galaxy.  From the $K$ and $B$ magnitudes (LEDA
Lyon-Meudon Extragalactic Database, http:// leda.univ-lyon1.fr/ and Fioc \&
Rocca-Volmerange 1999), the distances (van den Bergh 2000), the contribution
in near infrared light of young stars (Chiosi \& Vallenari 1996), a rough
estimate of the red giant population in each galaxy can be computed. A
typical luminosity of 100 L$_\odot$ for a cool giant is taken in computing
the number of red giants per galaxy.  The number of symbiotic stars can be
taken to be 0.5\% of the total number of red giants. Only a fraction of them
can be observed in the ``active'' phase, i.e. with a typical spectrum
containing bright emission line. From Table 1 it can be noted that the
number of expected symbiotic stars is quite dependent on the mass of the
galaxy.  For several irregular galaxies the number of symbiotic stars
estimated in this way is not reliable due the greater amount of $B$ light
respect to $K$ light.

\end{document}